\documentclass[%
preprint,
onecolumn,
superscriptaddress,
 amsmath,amssymb,
 aps, pre,
]{revtex4-2}

\usepackage{graphicx}
\usepackage{stmaryrd}%
\usepackage{xr}%
\usepackage{xcolor}
\usepackage{dcolumn}
\usepackage{bm}
\usepackage[mathlines]{lineno}

\graphicspath{{figures/}}

\newcommand{\change}[1]{#1}
\newcommand{\changepre}[1]{#1}
\newcommand{\changem}[1]{#1}

\begin{document}

\title{Multifractality in critical neural field dynamics}

\author{Merlin Dumeur}%
\affiliation{CEA, DRF, Joliot, NeuroSpin, Paris-Saclay University}
\affiliation{Inria MIND team, Paris-Saclay University}
\affiliation{Department of Neuroscience and Biomedical Engineering, Aalto University}

\author{Sheng H. \surname{Wang}}%
\affiliation{CEA, DRF, Joliot, NeuroSpin, Paris-Saclay University}
\affiliation{Inria MIND team, Paris-Saclay University}
\affiliation{Department of Neuroscience and Biomedical Engineering, Aalto University}
\affiliation{Neuroscience Center, Helsinki Institute of Life Science, Helsinki University}

\author{J. Matias \surname{Palva}}%
\thanks{J.M.P. and P.C. contributed equally to this work}
\affiliation{Department of Neuroscience and Biomedical Engineering, Aalto University}
\affiliation{Neuroscience Center, Helsinki Institute of Life Science, Helsinki University}
\affiliation{School of Psychology \& Neuroscience, University of Glasgow}

\author{Philippe Ciuciu}%
 \thanks{J.M.P. and P.C. contributed equally to this work}
\affiliation{CEA, DRF, Joliot, NeuroSpin, Paris-Saclay University}
\affiliation{Inria MIND team, Paris-Saclay University}
 

\date{\today}

\begin{abstract}

The brain criticality hypothesis has largely only characterized brain dynamics \change{in terms of their self-similarity}, although experimental evidence suggests that the brain exhibits significant multifractality.
To understand how multifractality may emerge in critical-like systems \change{modeling neuronal activity}, we used a \change{neural field} model exhibiting \change{neural oscillations and a critical phase transition}.
We find that multifractality emerges near a synchronization phase transition\change{, and that the pattern of variation of multifractality changes when placing the model at a different phase transition.}
These findings show \change{that} multifractality in temporal dynamics \change{emerges near criticality} in neural fields,
providing a \change{formal basis} for interpreting multifractality in brain recordings.


\end{abstract}

\maketitle



\section{\label{sec:intro} Introduction}

Although the scale invariance of critical systems in thermodynamic equilibrium is well established~\cite{Kardar2007StatisticalFields}, dynamic systems far from equilibrium and operating near a critical point also display scale invariance~\cite{Henkel2010Non-EquilibriumTransitions, Tauber2017}. In the brain, scale invariance is usually characterized by using avalanches~\cite{Beggs2003NeuronalCircuits} where both spatial and temporal scaling can be quantified, and by self-similarity analysis of brain oscillations and behavior~\cite{Hansen2001, Palva2013NeuronalLaws} when only temporal information is available.


For a time series, scale invariance can be expressed as the power-law scaling of the $q$-th (raw) absolute moments \changepre{of a statistical representation $T(j, k)$ of the signal as a function of temporal scale $j$ evaluated at discrete time points $k$, whose power-law exponent is the scaling function $\zeta(q)$}:
${\langle |T(j, k)|^q \rangle_k \propto 2^{j\zeta(q)}}$.
Self-similarity is characterized by the generalized Hurst exponent $H$~\cite{Mandelbrot1968FractionalApplications}, which is measured from $q=2$, corresponding to the power spectrum~\cite{Heneghan2000EstablishingProcesses}.
Although monofractal time series have a single parameter $H$ determining the scaling of every moment: $\zeta(q)=qH$, multifractal signals exhibit different scaling exponents at different statistical moments ~\cite{Stanley1988MultifractalChemistry},
as can be found in the field of physics of fully developed turbulence~\cite{Meneveau1991TheDissipation} as well as in financial time series analysis~\cite{Jiang2019MultifractalReview}.
In this case, the application of multifractal analysis (MFA) is warranted~\cite{Jaffard2004} to fully characterize $\zeta(q)$. 
For brain dynamics analysis, recent findings showed that self-similar infraslow (0.01-1 Hz) fluctuations of brain activity and behavioral performance~\cite{Monto2008VeryHumans, Palva2013NeuronalLaws, He2014Scale-freeFuture, ciuciu2014interplay}, exhibit significant multifractality~\cite{ciuciu2012scale, LaRocca2018}.

The brain is a large system that exhibits complex patterns such as spontaneous oscillations~\cite{Buzsaki2006}, whose dynamics have been suggested to arise from the brain operating near a critical point of a phase transition far from equilibrium,~\cite{Wright1985State-changesSteady-states, Breakspear2006AAnalysis} between synchronous and asynchronous phases~\cite{Scarpetta2014AlternationAttractors, Santo2018, Wang2023Critical-likeTransition, Fusca2023}.
Neuroscience research has used computational modeling to elucidate the mechanisms underlying the emergence of critical dynamics in brain-like systems~\cite{Zimmern2020}.
Thus, under the brain criticality hypothesis, the self-similarity in brain activity recordings is understood to originate from the brain's critical nature.

Multifractal analysis naturally extends self-similarity analysis; thus, the multifractal properties of neural time series are expected to be linked to brain criticality.
The analysis of the temporal dynamics of brain criticality models has remained largely restricted to measures of self-similarity such as \change{detrended fluctuation analysis}~(DFA)~\cite{Wang2023Critical-likeTransition}, without considering potential multifractality. Thus, the relationship between multifractal scaling and brain criticality~\cite{Zilber2014ERFParadigm, Alamian2022AlteredMagnetoencephalography} has remained speculative and has a limited basis in theory or generative models.

Here, we investigate the emergence of multifractal scaling behavior and its relationship with self-similarity by analyzing simulations of a field model of neuronal population activity, known as a 'neural mass model', which exhibits a critical point in the transition between synchronous and asynchronous phases~\cite{Santo2018}.

We report the scaling properties of fluctuations in both the envelope of the modeled oscillation and the low-frequency domain.
The model displays scale invariance across a range of temporal scales and, more importantly, a divergence of multifractal exponents near the critical point of the phase transition.
Furthermore, simulations indicate that multifractality is a consequence of finite-size effects and disappears in the thermodynamic limit.


\section{Model and multifractal formalism}

\subsection{Landau-Ginzburg theory for cortical dynamics}

The model simulated is derived from the Landau-Ginzburg field theory applied to the Wilson-Cowan neural mass model, as first proposed in~\cite{Santo2018}:
\begin{equation}
    \label{eq:model}
    \begin{cases}
        \dot{\rho} &= (R - a)\rho + b \rho ^2 - \rho ^3 + h + D \nabla^2 \rho + \sigma \sqrt{\rho}\eta \\
        \dot{R} &= \frac{1}{\tau_R} \left( \xi - R \right) - \frac{1}{\tau_D} R \rho ;.
    \end{cases}
\end{equation}

\begin{figure*}[t]
    \centering
    \includegraphics[width=\linewidth]{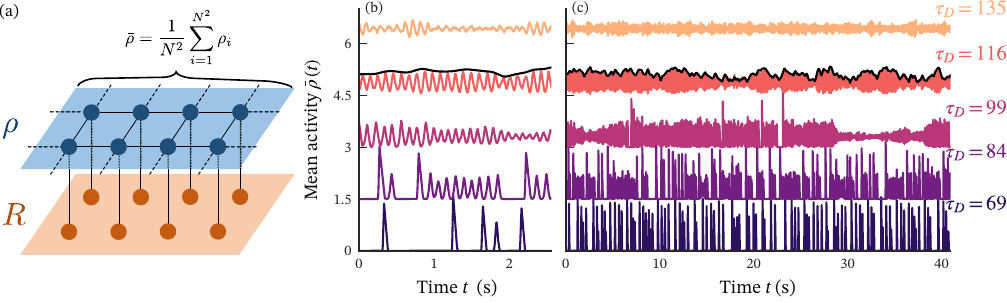}
    \caption{Schema of the discretized model (a) in blue, the excitatory activity field $\rho$; in orange, the resource field $R$.
    \change{Spatial average $\bar{\rho}$ of the activity field from model simulations, observed at the minimal (b) at maximal (c) characteristic temporal scales used in subsequent multifractal analysis. The values of the control parameter $\tau_D$ vary from subcritical (bottom) to supercritical (top) through the critical point around $\tau_D \approx 99$. Time series are offset on the y-axis by increasing multiples of 1.5 for legibility. The black line represents the extracted envelope of the oscillations for a single time series ($\tau_D=116$).}
}
    \label{fig:model}
\end{figure*}

The activity $\rho$ corresponds to the excitatory neural activity term in the Wilson-Cowan equations, of which the third-order Taylor expansion has been retained, revealing the simplest formulation of a differential equation with a first-order phase transition.
\changepre{The ``resources'' term $R$ acts on $\rho$ as the opposite of inhibitory neural activity ($\mathrm{d}\rho/\mathrm{d}R>0$), yet its effective action is still to reduce activity based on recent firing activity.}
Since it evolves at a much slower time scale than $\rho$ ($\tau_R \gg 1, \ \tau_D \gg 1)$, it can be understood as the quasi self-organized control parameter of the excitatory field equation.
Together, these coupled field equations reproduce the self-organized critical oscillation behavior~\cite{Buendia2020}, where neural oscillations emerge in a synchronization phase transition, concurrently with avalanches displaying power-law scaling. We investigate the phase transition by varying $\tau_D$ as our control parameter, which in the mean-field approximation controls the bifurcation from a limit cycle to bistable dynamics.

Fig.~\ref{fig:model} (b-c) presents the transition from a \change{subcritical, absorbing} down state of 
\change{zero activity interrupted by} intermittent synchronous activity ($\tau_D\leqslant80$), to a \change{supercritical} up state containing oscillations ($\tau_D>100$) showing continuous asynchronous activity. In between the down phase and the critical point lies a bistable phase, where the dynamics of the Wilson-Cowan model alternates between synchronous spiking and asynchronous oscillations~\cite{AlvankarGolpayegan2023BistabilityModel}.

Simulations of the model were carried out using a discrete version of the equations, approximating the fields with 2D square lattices, with edges of size $N$ and toroidal boundary conditions~(cf. Fig.~\ref{fig:model} (a)).

\subsection{Wavelet $p$-leader multifractal formalism}

\change{
Multifractal analysis characterizes the fluctuations of the pointwise regularity of scale-invariant processes. The multifractal spectrum associates to a given regularity $h$ the fractal dimension $D(h)$ of the set of time points with pointwise regularity $h$. Qualitatively, a multifractal process will present local fluctuations in regularity; thus, multifractality may be understood qualitatively as "burstiness": Where the regularity of the process is locally high, the signal is smooth, and where the regularity is locally low, the signal is very irregular.
A scale-invariant process which is monofractal thus does not show any variation in pointwise regularity, and will not display the bursts of activity characteristic of multifractal processes.
}

\change{
It is necessary to choose a multifractal formalism, which implies a choice of the specific notion of \textit{pointwise regularity} for the determination of the multifractal spectrum. We chose the recently introduced wavelet $p$-leader formalism~\cite{Jaffard2016P-exponentRegularity}, which relies on the pointwise $p$-exponent~\cite{Jaffard2005WaveletExponents} notion of regularity.
}

In this contribution, we rely on computing the wavelet $p$-leaders $\ell^{(p)}_{j, k}$, which are derived from the discrete wavelet coefficients $d_X(j, k)$ of the time series being analyzed.
Noting $\lambda_{j, k} = \left[ 2^j k, 2^j(k + 1) \right)$ the dyadic interval associated with the temporal scale scale $j$ and  temporal shift $k$ and $3\lambda_{j, k} = \lambda_{j, k-1} \cup \lambda_{j, k} \cup \lambda_{j, k+1}$ the dyadic interval centered around $\lambda_{j, k}$ with three times the width, then the wavelet $p$-leaders are defined as:
\begin{equation}
\ell^{(p)}_{j, k} := \left( \sum _{1\leq j'\leq j} \sum_{k\prime \in 3\lambda_{j, k}} \left| d_X(j', k') \right|^p 2^{j-j^{\prime }} \right)^{1/p} \,.
\end{equation}
Given the wavelet $p$-leader description of a scale-invariant time series, the $p$-leader scaling function $\zeta^{(p)}(q)$ characterizes the scaling behavior of its moments between the temporal scales $j_1 < j_2$:
\begin{equation*}
\mathbb{E}_k\left[ (\ell_{\lambda_{j, k}}^{(p)})^q \right] \propto 2^{j\zeta^{(p)}(q)}, \quad j\in [j_1, j_2]\,.
\end{equation*}

\noindent Taking the Taylor expansion of $\zeta^{(p)}(q)$ around zero:
\begin{equation*}
    \zeta^{(p)}(q) = q \ c_1^{(p)} + \frac{q^2}{2} c_2^{(p)} + \frac{q^3}{6} c_3^{(p)} + \sum_{k=4}^{+\infty} \frac{q^k}{k!} c_k^{(p)} \,,
\end{equation*}
where the $p$-leader log cumulants $\bigl( c_m^{(p)}\bigr)_m$ characterize the multifractal spectrum: $c_1$ corresponds to the mode of the spectrum and measures self-similarity; $c_2$ corresponds to the half-width of the spectrum and measures multifractality. They are determined by linear regression of the cumulant scaling functions $C_m^{(p)}(j)$ over the scale interval $[j_1, j_2]$. These functions are defined as the $m$th-order cumulants of the log of the wavelet $p$-leaders, and scale as power laws with exponents $c_m^{(p)}$:
\begin{equation*}
    C^{(p)}_m(j) = \mathrm{Cumulant}_m \left[ \left( \log_2 \ell^{(p)}_{j, k} \right)_k \right] = j c_m^{(p)} + K_{m}\,.
\end{equation*}


A process is said to be multifractal if and only if $c_2^{(p)} < 0$. Otherwise, the signal is monofractal and all its temporal scale-invariant properties are described by $c_1^{(p)}$. In this work, we restrict our analysis to the second-order cumulants ($m \leqslant 2$), as this is the only order required to demonstrate multifractality.

In the following, $p=2$ is fixed for comparability with the results of DFA; for the sake of clarity, the exponents $(p)$ are omitted from the subsequent notation.

\section{Results}

We simulated the model using the method proposed in~\cite{Dornic2005} and further improved in~\cite{Weissmann2018SimulationNoise}, which was implemented in Python and accelerated using a GPU.
\change{The length of the simulations is of the order of $3\cdot10^{6}$ samples, which means the temporal scales $j=12$ to $j=16$ correspond to intervals of size $5.12\cdot 10^3$ to $8.192\cdot 10^4$ samples.}

Except where specified, the simulation parameters were set to default values as follows: $\xi=2.47, \tau_D=100, \tau_R=1000, a = 1, b = 1.5, h = 10^{-7}, D = \Delta_x = \sigma = 1, \Delta_t = 0.01, N=128$. These correspond to the critical point of the phase transition investigated in~\cite{Santo2018}.
\changem{From a modeling perspective, the choice of timescale is essentially arbitrary. The model's relevance depends on its ability to correctly capture the essential properties of interest in brain dynamics.
For comparability with the wider literature, results are reported here using the timescale $1s = 2 \times 10^5 \Delta t$. This places the model in the range of dynamics where alpha band oscillations can be modeled using the same short-term plasticity rationale~\cite{Zonca2021EmergenceDynamics} that motivated the expression of $\dot{R}$ in~\eqref{eq:model}.
Therefore, the frequency associated with the scale $j$ is $f_j \approx \frac{3}{4} 2^ {11-j}$, and hence these time scales evolve as the inverse of the logarithm of the frequency.} 

The presence of a rare metastable steady state in the subcritical regime perturbs the proper estimation of the log-cumulants $c_1$ and $c_2$ in the low-frequency case, and was addressed by removing segments of the signal using a segmentation method described in the supplemental material (Supp.~Figure~1)~\cite{supplemental}.

Multifractal analysis was performed using an implementation of the wavelet p-leader framework \footnote{Made available at https://github.com/neurospin/pymultifracs}. The analysis wavelet is the Daubechies wavelet with 3 vanishing moments. Wavelet $p$-leader based analysis has the additional requirement that the wavelet scaling function should be positive $\eta(p)>0$. This is achieved by fractionally integrating the time series with coefficients $\gamma_{LF}=1.5$, and $\gamma_{Env}=1$, for the low frequency and envelope modalities, respectively.

\subsection{Scale invariance in the activity temporal trace}

\begin{figure*}[ht]
    \centering
    \includegraphics[width=\linewidth]{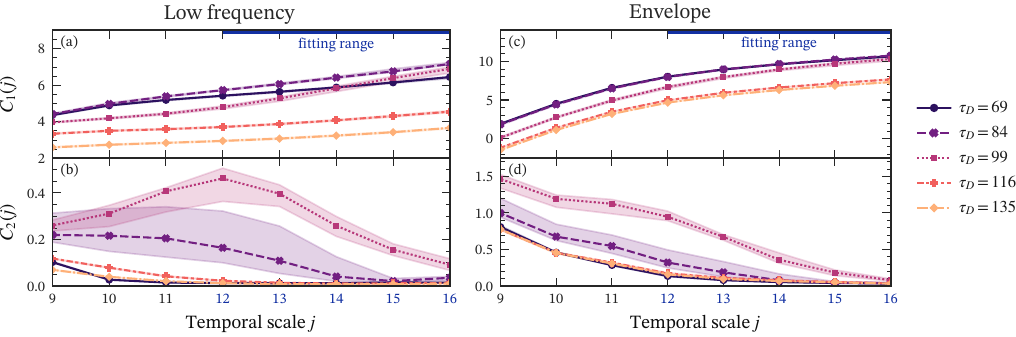}
    \caption{Cumulant scaling functions $C_1(j)$ (top row) and $C_2(j)$ (bottom row) averaged over 20 simulations, for the low-frequency domain (a-b) and oscillation envelope (c-d), for five $\tau_D$ values introduced in Fig.~\ref{fig:model}.
    Shaded area indicates 90\% confidence interval over 20 simulations.
    In blue: range of scales $[j_1, j_2]$ over which the log-cumulants were estimated.
    The characteristic temporal scale of the oscillation is $j=7.2$.}
    \label{fig:cumulants}
\end{figure*}

A necessary prerequisite for scale-invariance analysis is the presence of scale invariance in the data.
To meaningfully assess multifractality, we investigate the presence of scale invariance in the simulated time series and fit our estimation of the log-cumulants within the range of temporal scales over which the time series exhibits scale invariance. Under the wavelet $p$-leader formalism, scale-invariant time series have linear cumulant scaling functions $C_m(j)$.

The analysis is performed on the spatially coarse-grained excitatory activity process $\bar{\rho}(t) = \frac{1}{N^2}\sum_{i=1}^{N^2} \rho_i$.
We report scale invariance in the lower frequencies of the broadband signal, which is characteristic of critical dynamic models and which we term \textit{low frequency}.
The simulations are further filtered with a \change{complex} Morlet wavelet, which has a central frequency equal to the characteristic frequency of the oscillations ($\omega_0=0.0046 \mathrm{Hz}$) and a shape parameter $\omega = 13$. \change{Multifractal analysis applied to the absolute value of the complex filtered signal is labeled here \textit{envelope}, as it represents the envelope (i.e., instantaneous amplitude) of the filtered oscillatory signal.}

Fig.~\ref{fig:cumulants} shows cumulant scaling functions estimated from simulations. The five simulations depicted illustrate how the model behaves in terms of scaling functions $(C_1(j),C_2(j))_{j}$ with varying $\tau_D$.
The $C_m(j)$ show a crossover at the scale $j=12$, below which the scaling functions are dependent on characteristic effects from the evolution of the inhibitory field. Above that threshold, the scaling functions are linear, which indicates scale invariance of the time series; thus, the multifractal formalism is applicable.

We proceed with the multifractal analysis across the $j\in \llbracket 12, 16 \rrbracket$ range for all simulations and for both modalities.

\subsection{Multifractality near criticality}

\begin{figure*}[ht]
    \centering
    \includegraphics[width=\linewidth]{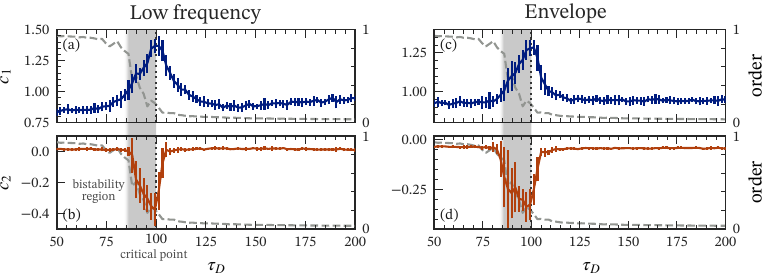}
    \caption{Continuous lines: log-cumulants $c_1$ (top row) and $c_2$ (bottom row) across the phase transition for the low frequency domain (a-b) and the envelope (c-d). Error bars show the standard deviation over 20 simulations. Dashed grey lines show model order (Kuramoto synchronization parameter $K$). Vertical dotted lines indicate the true critical point; the shaded area delimits the bistability region.
}
    \label{fig:critexp}
\end{figure*}

To assess the relationship between multifractality and the critical dynamics of the system, we now examine how the phase transition from the down state to the up state affects the log-cumulants $(c_m)_m$.
Varying the resource depletion rate characteristic time scale $\tau_D$, which controls the transition between limit cycle and bistable behavior in the mean field approximation of the model,  \change{the Kuramoto order parameter $K$ (Fig.~\ref{fig:critexp}, dashed gray line) characterizes the synchronization phase transition: the region of bistability begins approximately where $K$ starts to drops.}

Fig.~\ref{fig:critexp} presents $c_m$ as a function of the control parameter: $c_1$ measures self-similarity and $c_2$ measures multifractality.
Relating $(c_m)_m$ to the Kuramoto synchronization order parameter $K$ shows that, in the low-frequency fluctuations, $(c_m)_m$ take extreme values \change{near the critical point of the phase transition (dotted vertical line)}. Multifractality ($c_2 < 0$) is present near the critical point and extends all the way through the bistable region (shaded area), but is absent from the rest of the parameter space ($c_2 = 0$). \changepre{The degrees of multifractality and self-similarity, measured by $|c_2|$ and $c_1$, respectively, are highest at the critical point.}

Similarly, for the oscillation envelope, the extremal values of $c_m$ fall on the critical point. However, in contrast to the low-frequency case where $c_2=0$ far from the critical point, there is a constant \change{maximal} $c_2 < 0$ in the envelope estimates.

The fractional integration increases $c_1$ by the value of $\gamma$ and $\gamma_{LF} - \gamma_{Env} = 0.5$. However, the values of $c_1$ measured for the oscillation envelope are closer to the $c_1$ measured for the low-frequency component than would be expected from fractional integration. Therefore, the oscillation envelope is overall more regular than the low-frequency component.

\subsection{Different multifractal scaling near criticality}

\begin{figure*}[ht]
    \centering
    \includegraphics[width=\linewidth]{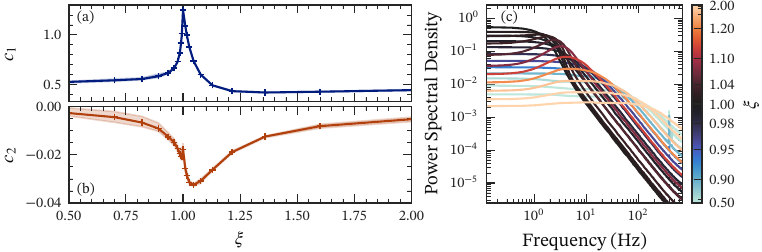}
    \caption{\change{Second order non-absorbing phase transition. Multifractal exponents $c_1$ (a) and $c_2$ (b) as a function of the parameter $\xi$, averaged across 10 simulations. The shaded area indicates the range of extreme observations. Power spectral density of the time series, with normalized energy (c); the color bar shows the values of $\xi$ used.}}
    \label{fig:secondorder}
\end{figure*}

\change{It is unclear, based on the above results, whether multifractality and self-similarity are essentially linked at criticality, and whether multifractality is a relatively trivial feature of highly self-similar critical dynamical systems. In this section, we place the model in a simple non-absorbing second-order phase transition by selecting the following parameters: $\tau_D = 1000, \sigma=0.1, h=10^{-6}, b=0$, and analyze the resulting patterns for $c_1$ and $c_2$ across the phase transition from a quiet phase with activity close to zero, to an always-active phase with no oscillations.}

\change{Fig.~\ref{fig:secondorder} presents the $c_1, c_2$ exponents as the baseline resources $\xi$ go from $0.5$ to $2.0$. Here $c_1$ has a visible peak at the critical point (just above $\xi = 1$), as in Fig.~\ref{fig:critexp}. In contrast, the $c_2$ exponent remains close to zero, denoting smaller levels of multifractality than in Fig.~\ref{fig:critexp}. Furthermore, $c_2$ is minimal around $\xi=1.1$, which is firmly in the supercritical domain, in contrast to previous results that showed a minimal $c_2$ at the critical point and overall increased multifractality in the subcritical domain.}

\section{Discussion}

Both the low-frequency fluctuations and the oscillation envelope in the model exhibit multifractality, with the peak occurring at the critical point. This shows that they both behave as critical exponents of the same phase transition, even though they are measured on distinct elements of the time series. This finding supports the empirical use of the self-similarity of the oscillation envelope as \change{a relevant metric for assessing the presence of criticality in brain activity recordings and motivates the application of multifractal analysis to the oscillation envelope of neuro physiological recordings.}

\change{
The bistable (subcritical) region of the model displays multifractality even at a distance from the critical point. Following the Kuramoto synchronization parameter, which is an order parameter of the critical phase transition, we understand the observed multifractality to reflect the near-critical dynamics of the specific bistable phase. The results presented for the second-order phase transition (Fig.~\ref{fig:secondorder}) clarify that multifractality is not specifically tied to bistability or to the mechanisms of first-order phase transitions, in general. However, whether the relationship between self-similarity and multifractality depends on the nature of the phase transition remains an open question. An approach based on large deviation theory could, in principle, provide theoretical multifractal spectra for systems far from equilibrium~\cite{Zohar1999LargeMultifractals, Harris2007FluctuationDynamics, Touchette2013LargeSystems}.
}

The results presented here may be easily extended to other neural field theories that implement self-organized critical oscillations~\cite{Buendia2020}, namely models that have two coupled fields: the first field showing a first-order phase transition~(equivalent to $\rho$), coupled to another slower evolving field that controls the phase transition~(equivalent to $R$). This represents an abstraction of brain activity that will require adjusting to real-life constraints and parameters; however, the results are expected to be qualitatively similar.

The presence of temporal multifractality in this model supports the empirical findings showing that multifractality is physiologically relevant either in the low-frequency fluctuation case~\cite{LaRocca2018}, or in the oscillation envelope~\cite{Catrambone2021FunctionalDomain}. In general, the intermittency of brain oscillatory processes has been described as physiologically relevant~\cite{Jones2016WhenMeaning}, but remains unclear in characterizations strictly limited to self-similarity.

\change{In the context of brain criticality research, the results presented here motivate the use of multifractal analysis as a tool to understand the organization of brain activity. The fact that self-similarity and multifractality are linked to criticality in the model may serve as a starting point for formulating hypotheses about the brain's dynamics, which can be tested to validate the modeling approach.}
\change{As an example for a clinical setting: Does the relationship between self-similarity and multifractality differ significantly between control and clinical subjects? This would indicate a difference in the nature of the underlying phase transition that may not be explained by purely linear dynamical effects in the brain.
Multifractality may also be assessed to identify anomalous behavior in large datasets. Significant deviations from the population baseline may reveal a new signal that current tools fail to capture.}
\change{Importantly, the presence of apparent multifractality in the oscillation envelope far from the critical point suggests that the envelope extraction process may induce a small degree of spurious multifractality, which should then be addressed when performing statistical tests, for instance by using permutation-based surrogates.}

In conclusion, temporal multifractality emerges from self-organized critical oscillation theory and provides a plausible explanation for the previously tentative link between self-organized brain criticality and empirical multifractality measured in cerebral time series.
~
\begin{acknowledgments}

This work is supported by the "ADI 2020" project funded by the IDEX Paris-Saclay, ANR-11-IDEX-0003-02 to M.D.;
by ANR 19-CE48-0002-04 DARLING  to M.D. and P.C.;
by Sigrid Juselius Foundation to M.D., S.W. (210527), and J.M.P.;
by the Instrumentarium Foundation to M.D.;
and by the Finnish Cultural Foundation postdoc fellowship (00220071) to S.W.
The simulations and analysis presented above were performed using computer resources within the Aalto University School of Science “Science-IT” project. 

\end{acknowledgments}

\appendix

\nocite{Truong2020SelectiveMethods, Wendt2008, Chhabra1989DirectTurbulence}

\bibliography{references}

\end{document}